# Paramagnetic tunneling state concept of the low-temperature magnetic anomalies of multicomponent insulating glasses

Alexander Borisenko[*] and Alexander Bakai

*National Science Center "Kharkiv Institute of Physics&Technology", 1 Akademichna str., 61108 Kharkiv, Ukraine*

(Dated: March 17, 2006)

A generalized tunneling model of multicomponent insulating glasses is formulated, considering tunneling states to be paramagnetic centers of the electronic hole type. The expression for magnetic field dependent contribution into the free energy is obtained. The derivation is made of the expression for the nonmonotonic magnetic field dependence of dielectric susceptibility, recently observed in amorphous $BaO-Al_2O_3-SiO_2$ in sub-Kelvin temperature range.

In the past few years a puzzling behavior of some multicomponent insulating glasses in magnetic field at sub-Kelvin temperatures was discovered. For example, the dielectric constant $\varepsilon$ of amorphous $BaO-Al_2O_3-SiO_2$ (BAS) in complex nonmonotonic manner depends on the magnetic field at $T \leq 1$ K [1, 2]. Moreover, a sharp kink in $\varepsilon$ for this material is observed, indicating a spin glass-type transition at $T=5.84$ mK [3]. Within the millikelvin temperature range BAS also demonstrates the pronounced dependence of spontaneous polarization echo amplitude on magnetic field [4]. Similar results are reported for another multicomponent glass BK7 [5]. This B-field dielectric response is strongly dependent on the magnitude of driving voltage [2, 5]. The heat capacity of these materials also depends on magnetic field in a nonmonotonic manner (see [6] and references therein).

It should be mentioned that the magnitudes of these anomalous effects don't simply scale with the concentration of magnetic impurities in the samples [5], thus allowing to rule out their direct impact. Neither these properties can be interpreted as a magnetoeffect, characteristic to nonlinear dielectrics, where the quadratic B-field dependence of $\varepsilon$ should be found [7].

The obvious conclusion from the above facts is that some glasses posses a subsystem that is susceptible to magnetic fields and that this subsystem is related to the structural features of multicomponent amorphous insulators.

At zero magnetic field the anomalous physical properties of vitreous insulators at $T \leq 1$ K are more or less successfully described by the tunneling model (see [8] and references therein).

In the simplest case a tunneling state (TS) can be considered as an effective particle confined in a double well (W) potential. In this case due to overlap of the ground state wavefunctions in the two wells the ground energy level splits into a doublet with a gap $E = \sqrt{h^2 + \Delta^2}$, $\Delta$ being the gap value for symmetric potential, $h$ being the difference of ground state energies in two wells neglecting tunneling. Any other states of this system, except this doublet, are neglected. The parameters $h$, $\Delta$ are commonly assumed to be random, obeying the phenomenological distribution:

$$P(h,\Delta) = P_0/\Delta, \quad E_{\min} \leq E(h,\Delta) \leq E_{\max}. \tag{1}$$

The phenomenological parameters of the distribution function (1) are the constant $P_0$, proportional to the volume density of TSs and the lower and upper energy cutoffs $E_{\min}$ and $E_{\max}$. Although one has no reason to believe that the distribution function has so simple form for all substances in the wide region of parameters and temperatures, it can be used as a trial, properly parametrized function.

It is natural to assume that TSs are responsible for the mentioned above puzzling magnetic properties too. The models elaborated up to now may be separated into "orbital" and "spin" group according to the assumed mechanism of TS coupling to magnetic field.

In the "orbital" models, tunneling of an electrically charged particle between potential minima can occur along different paths, so the presence of a magnetic field yields an Aharonov-Bohm phase and a change of energy eigenvalues. In a recent publication [7] a hat-shaped W potential was considered, with two minima in the azimutal direction along the rim. Within this model, the experimentally observed for BAS samples maximum in the real part of dielectric constant $(\varepsilon')$ at $B \approx 0.1\,\text{T}$ (data taken at the driving electric field amplitude $\approx 15\,\text{kV/m}$) requires one to assume the TS electric charge to be $q \sim 10^5\,|e|$, where $e$ is an elementary charge. The authors speculate the origin of such a large value of $q$ to result from the strong cooperative interaction between TSs at low temperatures, when the quasiparticles should be considered rather then the "bare" TSs.

Instead of a hat-like potential a multi-well potential may be considered. G. Jug [6] considers a 3-well potential with shallow minima, so that $\Delta$ is of the same order of magnitude as the single-well ground energy value. This gives an energy spectrum, consisting of the nondegenerate excited state and doubly degenerate (for the symmetric case) ground state. Magnetic field breaks the degeneracy of this doublet and opens a gap $\Delta E \propto \Delta \cdot \phi$, where $\phi$ is the

Aharonov-Bohm phase. Due to the large value of $\Delta$ taken, this model produces a good fit of heat capacity data for some glasses, assuming the product of the area encompassed by the tunneling particle times its charge in units of elementary charge to be of the order $10^2 \text{ Å}^2$. This model also gives the nonmonotonic B-field dependence of $\varepsilon$ [9].

The multi-path tunneling may be realized in a different way, considering interaction between pairs of TSs of certain relative orientation and closed tunneling sequences in this complex [10]. In the presence of magnetic field this system possesses two orbital quantum states with a linear B-field dependence of the energy gap between them. Under the experimental conditions, in the strongly nonlinear dynamic regime, the deduced dielectric susceptibility shows an oscillatory behavior, with an effective flux quantum of the order of $10^{-5} \hbar/e$. However, this model is formulated for the near-degenerate TSs $(h \ll \Delta)$ and, as we are concerned, it hasn't been developed for the case of general W potential asymmetry up to now.

In the models of "spin" group the intrinsic magnetic moment associated with the tunneling entity is considered. The model by Würger, Fleischmann and Enss [11] considers nuclear origin of spins of TSs. In the case of nonzero nucleus orbital moment its quadrupole moment is not zero and depends on the nucleus spin projection. The tunneling motion is then coupled to Zeeman energy due to the inhomogeneity of electric molecular field. This model seems to be adequate to explain the polarization echo experiments, but it fails to describe the B-field dependence of $\varepsilon$ even qualitatively [12].

In this paper we consider a magnetic field effect on TS, assuming it to be an electronic shell paramagnetic center, the paramagnetic tunneling state (PTS). The possible mechanisms of TS electronic paramagnetism origin in amorphous dielectrics may be different:

i) The presence of impurity paramagnetic ions (e.g. Fe) in the samples.

ii) The presence of unsaturated ("dangling") covalent bonds due to the break of local order [13] in the vicinity of W potential in chemically pure samples;

iii) The presence of localized electrons or holes due to chemical impurities with chemical valencies different from those of the host atoms.

It was argued [5] that mechanism (i) does not play a decisive role in the phenomena considered. And the fact that only multicomponent glasses have revealed the puzzling dielectric properties in magnetic field up to now brings us to the case (iii). As soon as all the major impurities (Al, B, Ba) present in amorphous $SiO_2$ samples under consideration possess chemical valencies smaller then that of Si, we shall concentrate on electronic hole localized on the oxygen ion (which is engaged into tunneling motion) in the vicinity of chemical impurity.

For the sake of simplicity we consider a single electronic hole with the total angular moment $\mathbf{J}^2 = \hbar^2 J(J+1)$, which results in a magnetic moment $\boldsymbol{\mu}_J = g_J \beta \mathbf{J}$, with the Bohr magneton $\beta$ and the Lande factor $g_J$. Its spin Hamiltonian consists from the term, which accounts for the interaction with external magnetic field (Zeeman energy) and (for $J \geq 1$) the quadrupole interaction with the gradient of inhomogeneous "crystal" field [14]. The quadrupole moment of the system results from the nonspherical charge distribution in the shell with nonzero orbital moment. For a symmetric electronic charge distribution $\rho(\mathbf{r})$ along the axis $\mathbf{u}$ its diagonal component reads:

$$Q_{uu} = \int d^3 r \left[ 3(\mathbf{r} \cdot \mathbf{u})^2 - r^2 \right] \rho(\mathbf{r}). \qquad (2)$$

Expression (2) may be rewritten in terms of the total angular moment projection $J_u = (\mathbf{J} \cdot \mathbf{u})/\hbar$ as follows [15]:

$$Q_{uu} = \frac{3 Q_J}{J(2J-1)} \left[ J_u^2 - \frac{1}{3} J(J+1) \right]. \qquad (3)$$

For the hole with the charge $+|e|$, mean square of the shell radius $\overline{r^2}$ and given angular quantum number $J = L \pm 1/2$ the value $Q_J$, which itself is often called a quadrupole moment, reads:

$$Q_J = -|e| \overline{r^2} \frac{2J-1}{2J+2}. \qquad (4)$$

Then in presence of external magnetic field $\mathbf{B} = B \mathbf{e}_z$ and the gradient of electric "crystal" field, in the simplest case described by the single diagonal term of the potential curvature $\varphi''(\mathbf{r}) \equiv \varphi''_{uu}(\mathbf{r}) = (\mathbf{u} \cdot \nabla)^2 \varphi(\mathbf{r})$, the spin Hamiltonian reads:

$$\hat{V} = g_J \beta \hat{J}_z B + \frac{\varphi''(\mathbf{r})}{4} \frac{Q_J \left[ 3 \hat{J}_u^2 - J(J+1) \right]}{J(2J-1)}. \qquad (5)$$

The spin Hamiltonian (5) has the same form as the one proposed in the model by Würger, Fleischmann and Enss [11] for nuclear spins.

In a general case, when both the magnetic field and the potential curvature are not zero and the axes $\mathbf{u}$ and $\mathbf{e}_z$ are not parallel, the Zeeman and quadrupole terms in Hamiltonian (5) can not be diagonalized simultaneously. For the estimation of quadrupole term in Hamiltonian (5) we take the value for the "crystal" electric field gradient in the minimum of "soft" W potential: $\varphi'' \sim 10^{18} \text{ V} \cdot \text{m}^{-2}$ and the value for the quadrupole moment of order of the square of $O^{2-}$ ion radius times the elementary charge: $|Q_J| \sim 10^{-39} \text{ Q} \cdot \text{m}^2$ to obtain $|V_Q| \sim 10^2 \text{ K}$. The Zeeman term reaches this value at the characteristic magnetic field strength $B \sim 100 \text{ T}$, well above the practically attainable limit. So, in this case one can treat the Zeeman energy as a small perturbation, using $\mathbf{u}$ as an appropriate quantization axis. This behavior of electronic holes is entirely different from that of nuclei, for which Zeeman and quadrupole terms of the spin Hamiltonian become comparable already at $B \sim 0.1\,\text{T}$ [12]. Now in the first order of perturbation theory we obtain the energy spectrum:

$$V_{J_u} = g_J \beta J_u B(\boldsymbol{e}_z \cdot \boldsymbol{u}) + \frac{\varphi''(r)}{4}\frac{Q_J[3J_u^2 - J(J+1)]}{J(2J-1)}. \quad (6)$$

From (6) one can see that that in the first order of perturbation theory the magnetic field, whenever is not orthogonal to the gradient of "crystal" electric field, lifts the degeneracy of the levels $\pm J_u$.

From (4) one can see that the quadrupole moment associated with the hole is negative and taking into account that the potential curvature $\varphi''(r_{\min})$ is positive in potential minimum, one finds that at zero magnetic field $J_u = \pm J$ in the ground state. From the estimations of quadrupole energy above one finds that levels with different $|J_u|$ are separated by the gaps of order of 100 K, thus allowing to consider only the doublet with $J_u = \pm J$ at sub-Kelvin temperatures. So, the single-well spin Hamiltonian can be treated in the effective spin-1/2 representation.

Now we address the case of two adjacent potential minima separated by the shallow barrier, transparent for the tunneling particle (Fig. 1).

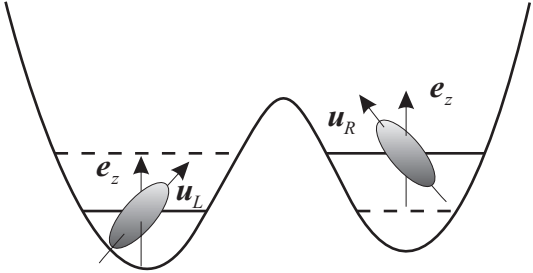

Fig. 1. W potential with non-parallel orientation of the hole quadrupoles in different wells.

Due to the topological disorder, characteristic to the glasses, the axes of the "crystal" field gradient in the adjacent wells may be non-parallel, leading to the non-conservation of the angular moment of tunneling particle. In this case, if the tunneling process takes place with a frequency $\Delta/\hbar$, where $\hbar$ is the Planck's constant, one can say that tunneling accompanied by the change $(J_u \leftrightarrow -J_u)$ and conservation $(J_u \leftrightarrow J_u)$ of spin projection takes place with frequencies $D/\hbar$ and $\delta/\hbar$ respectively [14], where

$$D = \Delta \cdot \sin\frac{\alpha}{2},\ \delta = \Delta \cdot \cos\frac{\alpha}{2} \quad (7)$$

and $\alpha \in [-\pi;\pi]$ is the angle between the axes $\boldsymbol{u}_L$ and $\boldsymbol{u}_R$ in the left- and right-hand wells respectively.

The single PTS Hilbert space is a product of subspaces of coordinate and spin states. In both subspaces the Pauli matrix representations are valid. Let the coordinate states ($|l\rangle$ and $|r\rangle$) correspond to the different eigenvalues of matrix $\sigma_z$, and the states with different signs of spin projection $J_u$ ($|u\rangle$ and $|d\rangle$) correspond to the different eigenvalues of matrix $\tau_z$.

The tunneling transition operator between states with equal spin projections is then represented by the Pauli matrix $\sigma_x$. The frequency of this tunneling motion is taken to be $\delta/\hbar$ (see (7)).

The tunneling transitions in which both the spin and coordinate states are changing simultaneously, are convenient to describe in terms of creation and annihilation operators:

$$\sigma^+ = (\sigma_x + i\sigma_y)/2\,;\ \sigma = (\sigma_x - i\sigma_y)/2$$
$$\tau^+ = (\tau_x + i\tau_y)/2\,;\ \tau = (\tau_x - i\tau_y)/2, \quad (8)$$

which act in the following way:

$$\sigma^+|l\rangle = |r\rangle\,;\ \sigma^+|r\rangle = 0\,;\ \sigma|l\rangle = 0\,;\ \sigma|r\rangle = |l\rangle\,;$$
$$\tau^+|u\rangle = |d\rangle\,;\ \tau^+|d\rangle = 0\,;\ \tau|u\rangle = 0\,;\ \tau|d\rangle = |u\rangle. \quad (9)$$

We take the frequency $D/\hbar$ (see (7)) for the tunneling transitions $|ru\rangle \leftrightarrow |ld\rangle$, governed by the operator $\sigma\tau^+ + \sigma^+\tau$. The tunneling transitions $|lu\rangle \leftrightarrow |rd\rangle$, governed by the operator $\sigma^+\tau^+ + \sigma\tau$, may be deduced from the above scheme by inversion of time $(t \to -t)$ and hence the frequency $-D/\hbar$ is associated with them.

Note that the coordinate and spin variables are considered to be independent, and hence $\sigma_i$ and $\tau_j$ commute with each other.

One should also take into account that Zeeman splittings of the ground energy levels in the two wells are different in general case: $u_L^Z = 2g_J\beta B(\boldsymbol{e}_z \cdot \boldsymbol{u}_L)$, $u_R^Z = 2g_J\beta B(\boldsymbol{e}_z \cdot \boldsymbol{u}_R)$, so it is convenient to introduce the new variables:

$$u_+^Z = \frac{u_R^Z + u_L^Z}{2}\,;\qquad u_-^Z = \frac{u_R^Z - u_L^Z}{2}. \quad (10)$$

In these notations, taking $h$ as the difference of ground state energy levels in the two wells, we can write down the single PTS Hamiltonian as follows:

$$\hat{H} = -1/2 \cdot \left(h\sigma_z + \delta\sigma_x + u_+^Z\tau_z + u_-^Z\tau_z\sigma_z + D\sigma_y\tau_y\right). \quad (11)$$

We assume direct coupling of the potential asymmetry $h$ to electric field $\boldsymbol{E}$ through the PTS intrinsic electric dipole moment $\boldsymbol{p}$:

$$h = h_0 + 2(\boldsymbol{p} \cdot \boldsymbol{E}). \quad (12)$$

The exact PTS Hamiltonian (11) is difficult to diagonalize algebraically, because its eigenvalues are not symmetrical with respect to zero energy level. For this reason we treat the simplified Hamiltonian with $u_-^Z = 0$, corresponding to the situation when the external magnetic field is directed along the bisector line of the angle, formed by the axes $u_L$ and $u_R$.

This simplified Hamiltonian has the form:

$$\hat{H} = -1/2 \cdot \left( h\sigma_z + \delta\sigma_x + u^Z \tau_z + D\sigma_y \tau_y \right), \quad (13)$$

where we denote $u^Z \equiv u_+^Z$ for simplicity.

Writing $\hat{H}$ in the basis $\{|lu\rangle, |ru\rangle, |ld\rangle, |rd\rangle\}$ and diagonalizing the resulting 4x4 matrix, we obtain the energy spectrum:

$$E_{S,A} = \mp \frac{1}{2}\sqrt{G_+^2 + D^2}; \quad E_{1,2} = \pm \frac{1}{2}\sqrt{G_-^2 + D^2} \quad (14)$$

and the normalized eigenvectors:

$$\psi_S = \frac{1}{2}\begin{pmatrix}1\\1\\1\\1\end{pmatrix}, \quad \psi_A = \frac{1}{2}\begin{pmatrix}-1\\1\\1\\-1\end{pmatrix}, \quad \psi_{1,2} = \frac{1}{2}\begin{pmatrix}-1\\\pm 1\\\mp 1\\1\end{pmatrix}. \quad (15)$$

In eq. (14) the abbreviations $G_\pm \equiv u^Z \pm \sqrt{h^2 + \delta^2}$ are used.

As expected, the fully symmetric state $\psi_S$ has the lowest energy, the fully antisymmetric state $\psi_A$ has the highest energy, while the mixed states $\psi_{1,2}$ form an internal doublet.

Note that nonzero value of the parameter $D$ prevents the crossover of energy levels $E_{1,2}$. This feature is essential in our model to describe the nonmonotonic magnetic field dependence of PTS contribution into dielectric susceptibility. By putting $u^Z = 0$ and taking into account (7), one obtains from (14) the result for conventional TS, with two doubly degenerate energy levels.

Using expression for the single PTS energy spectrum (14) and putting the Boltzmann's constant $k_B = 1$, one can obtain an expression for the single PTS free energy $f = -T \ln Sp \exp(-\hat{H}/T)$ in the explicit form:

$$f = -T \ln \left\{ 2\cosh\frac{\sqrt{G_+^2 + D^2}}{2T} + 2\cosh\frac{\sqrt{G_-^2 + D^2}}{2T} \right\}. \quad (16)$$

The thermodynamic variable of our special interest is a dielectric susceptibility. For the static case in the linear-response approximation we have:

$$\chi_{\alpha\beta}(\omega = 0) = -\frac{1}{V}\left(\frac{\partial^2 f}{\partial E_\alpha \partial E_\beta}\right), \quad (17)$$

where $E$ is an applied electric field, $V$ is the sample volume and $\alpha, \beta$ are the Cartesian indices.

To account for the PTS relaxation dynamics at low (on the energy gap scale) frequency $\omega$ of the harmonic external driving field we separate $\chi(\omega)$ into the resonant $\chi_{res}$ and relaxation $\chi_{rel}$ parts. This separation is natural if one considers the fact that the expression for PTS polarization $P_\alpha = -\frac{1}{V} \cdot \frac{\partial f}{\partial E_\alpha}$ contains both the expressions for statistical populations of energy levels and the effective magnitudes of dipole moments (different for the internal and external energy doublets). These latter are different from the absolute value of the PTS intrinsic electric dipole moment $p$ due to the tunneling overlap of states with different signs of polarization. Thus the expression for $\chi_{res}$ will contain the yield from differentiation of the values of the effective dipole moments at constant levels' populations and the expression for $\chi_{rel}$ will contain the yield from the levels' populations differentiation at constant values of the effective dipole moments. Then in the frame of $\tau$-approximation we obtain the next formula for $\chi(\omega)$:

$$\chi(\omega) = \chi_{res} + \frac{\chi_{rel}}{1 + i\omega\tau}, \quad (18)$$

where $\tau$ is a PTS relaxation time.

We use the next expression for the phonon-mediated PTS relaxation time:

$$\tau(\Delta, T) = \left( \gamma \cdot \Delta^2 \cdot T \right)^{-1}, \quad (19)$$

where $\gamma$ is a material-dependent constant. Expression (19) is a natural generalization of that for conventional TS [16], assuming the energy gaps to be less then $T$. In this assumption the energy gaps (and hence the magnetic field) do not enter the expression for relaxation time, so (19) should be considered as a low B-field limit. The tunneling parameter $\Delta^2 = \delta^2 + D^2$ (see (7)) accounts for the both types of tunneling processes considered in our model.

The explicit form of expressions for $\chi_{res}$ and $\chi_{rel}$ is a bit cumbersome to display here. It comes out that all the terms in eq. (17) proportional to $1/T$ concern the relaxation component, while the rest contribute to the resonant part.

Both $\chi_{res}$ and $\chi_{rel}$ contain the terms which depend nonmonotonically on Zeeman splitting, for nonzero $\sqrt{h^2 + \delta^2}$. We denote them $\chi_{res}^{nm}$ and $\chi_{rel}^{nm}$ respectively:

$$\chi_{res}^{nm} \propto 2\sinh\frac{\sqrt{G_-^2 + D^2}}{2T} \frac{h^2 D^2}{(h^2 + \delta^2)[G_-^2 + D^2]^{3/2}} \quad (20)$$

$$\chi_{rel}^{nm} \propto \cosh\frac{\sqrt{G_-^2 + D^2}}{2T} \frac{G_-^2 h^2}{(h^2 + \delta^2)[G_-^2 + D^2]T} \quad (21)$$

At $u^Z = \sqrt{h^2+\delta^2}$ $\chi_{res}^{nm}$ has a maximum, while $\chi_{rel}^{nm}$ has a minimum, which cancel each other for $D \leq T$. The widths of these nonmonotonic regions are proportional to $D$.

Numerical calculations of PTS dielectric susceptibility show that the extrema shapes and positions depend on the assumed distribution of PTS parameters.

For illustration of our results in Figs. 2, 3 we present experimental data for BAS [2] and BK7 [5], together with our linear-response calculations for the PTS contribution into ac dielectric constant (17 - 18), integrated over the trial distribution function (1), assuming the dipole moments to have a fixed absolute value $p_0$. The angle between axes $\boldsymbol{u}_L$ and $\boldsymbol{u}_R$ is taken to be constant. As paramagnetic center an electronic hole in the state $^2P_{3/2}$ is considered, corresponding to the ground state of O$^-$ ion. The frequency of electric driving field $\omega$=1 kHz is taken.

Table 1. The values of fitting parameters.

| $\dfrac{4\pi P_0 p_0^2}{3\varepsilon_0}$ | $E_{min}$, K | $E_{max}$, K | $\gamma$, K$^{-3}$s$^{-1}$ | $\alpha$ | $J$ | $g_J$ |
|---|---|---|---|---|---|---|
| $3\cdot 10^{-4}$ | $10^{-6}$ | 10 | $5\cdot 10^7$ | 2.32 | 3/2 | 4/3 |

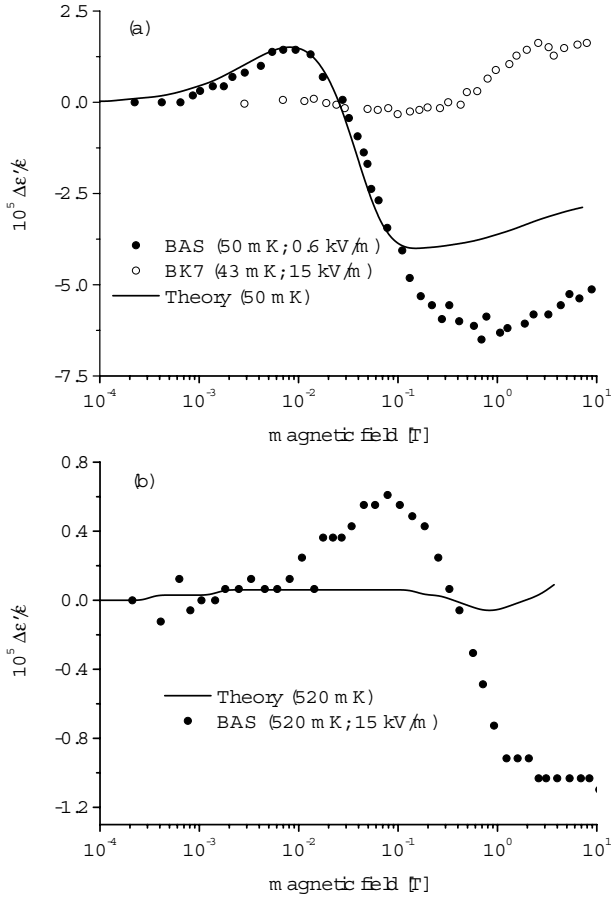

Fig. 2. The relative change of the real part of glass's dielectric constant vs. applied magnetic field. (a) Experimental data for BAS [2] and BK7 [5] and our linear-response fit at conditions as indicated. (b) Experimental data for BAS [2] and our linear-response fit at conditions as indicated.

The values of fitting parameters are given in Table 1, taking $\varepsilon_0$ as a vacuum electric constant.

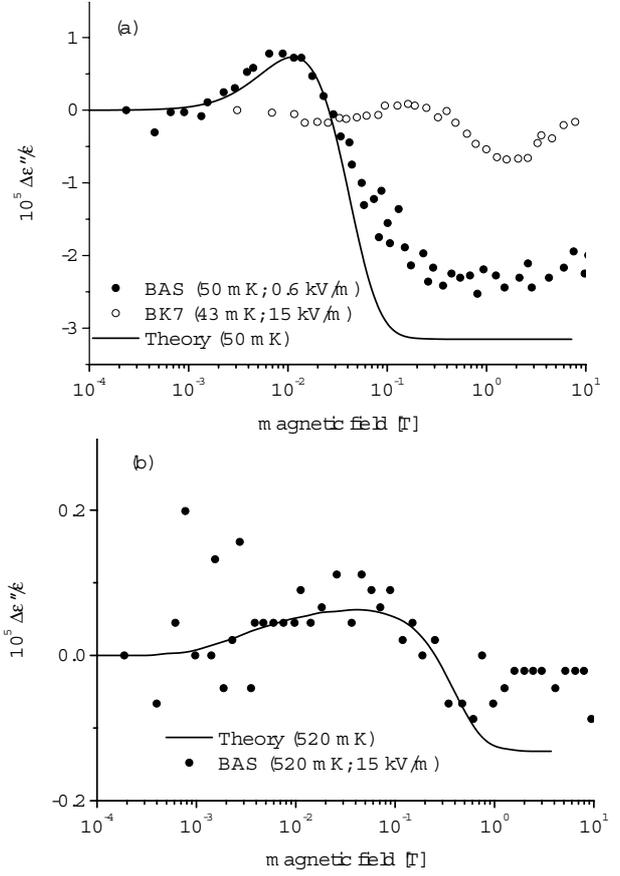

Fig. 3. The relative change of the imaginary part of glass's dielectric constant vs. applied magnetic field. (a) Experimental data for BAS [2] and BK7 [5] and our linear-response fit at conditions as indicated. (b) Experimental data for BAS [2] and our linear-response fit at conditions as indicated.

From the experimental data in Figs. 2a, 3a one can see that the two types of glasses respond to the magnetic field in a qualitatively different manner. On the base of our knowledge we can not provide an ultimate explanation of this difference. It looks like that there are much less (if any) PTSs in BK7 than in BAS. This difference can be attributed to the microscopic structure of these glasses.

On the base of the present theory the quantitative agreement with the data for BAS at $T$=50 mK (taken at low driving voltage $\approx$0.6 kV/m, which allows the validity of the linear-response approximation) can be achieved in the region of comparatively small magnetic fields $(B \leq 0.1\,\text{T})$. The discrepancy at higher magnetic fields can be possibly attributed to the use of the approximate expression for PTS relaxation time (eq. (19)) in our calculations.

For the more thorough check of our theory we also made calculations for $T$=520 mK (Figs. 2b, 3b). Unfortunately, the experimental data available at this temperature are taken at much higher magnitude of driving voltage $\approx$15 kV/m. At these conditions we were not able to achieve an agreement for the real part of dielectric constant, Fig. 2b, but the agreement is satisfactory for $B \leq 1$ T for the imaginary part (which seems to be much less sensitive to the magnitude of applied voltage) within the scatter of experimental data, Fig. 3b.

Comparing the experimental data at two different temperatures, one can conclude that the magnetic field values at which maxima of the real and imaginary parts occur, roughly scale with temperature, thus implying a broad distribution of PTS parameters $h$, $\Delta$.

In summary, we propose a PTS model for multicomponent amorphous insulators, which assumes that due to the local chemical disorder a tunneling entity may have holes in its electronic shell, and therefore to be a paramagnetic center. The tunneling motion leads to the non-conservation of magnetic moment due to the disorientation of the hole-associated quadrupole moment by the random "crystal" field. This feature gives rise to the nonmonotonic magnetic field dependence of the dielectric susceptibility of PTS ensemble. By the use of this theory we were able to obtain a semi-quantitative agreement with experimental data for amorphous $BaO-Al_2O_3-SiO_2$ [2].